\documentclass{iau} 
\usepackage{graphicx}
\usepackage{hyperref}

\title[The e-ASTROGAM gamma-ray space mission] 
{The e-ASTROGAM space mission: a major step forward for supernova physics}

\author[V. Tatischeff, R. Diehl \& A. De Angelis]   
{Vincent Tatischeff$^1$,
 Roland Diehl$^2$,
 \and Alessandro De Angelis$^3$, \\
 on behalf of the e-ASTROGAM Collaboration
 \thanks{See \url{http://eastrogam.iaps.inaf.it/}}
 }

\affiliation{$^1$CSNSM, IN2P3-CNRS/Univ. Paris-Sud, Universit\'e Paris-Saclay, F-91405 Orsay Campus, France \\ email: {\tt vincent.tatischeff@csnsm.in2p3.fr} \\[\affilskip]
$^2$Max Planck Institut f\"ur extraterrestrische Physik, D-85748 Garching, Germany \\ email: {\tt rod@mpe.mpg.de}\\[\affilskip]
$^3$INFN Padova, via Marzolo 8, 35141, Padova, Italy; LIP/IST, Av. Elias Garcia 14, 1000 Lisboa, Portugal
\\ email: {\tt alessandro.deangelis@pd.infn.it}}

\pubyear{2017}
\volume{xxx}  
\jname{Title of your IAU Symposium}
\editors{A.C. Editor, B.D. Editor \& C.E. Editor, eds.}
\begin{document}

\maketitle

\begin{abstract}
e-ASTROGAM is a gamma-ray observatory operating in a broad energy range, 0.15~MeV~--~3~GeV, recently proposed as the M5 Medium-size mission of the European Space Agency. It has the potential to revolutionize the astronomy of medium-energy gamma-rays by increasing the number of known sources in this domain by more than an order of magnitude and providing gamma-ray polarization information for many of these sources. In these proceedings, we discuss the expected capacity of the mission to study the physics of supernovae, both thermonuclear and core-collapse, as well as the origin of cosmic rays in SN shocks. 

\end{abstract}

\firstsection 
\section{Introduction}

\begin{figure}
\begin{center}
\begin{tabular}{c}
\begin{minipage}{0.15\linewidth}
\includegraphics[scale=0.18]{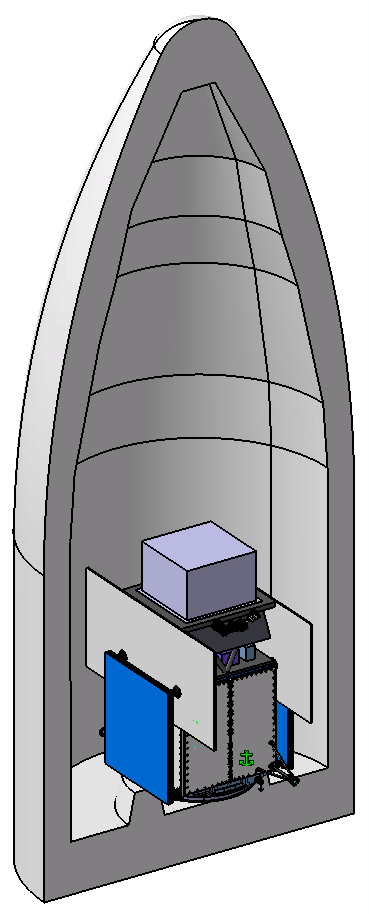}
\end{minipage}
\begin{minipage}{0.45\linewidth}
\includegraphics[scale=0.148]{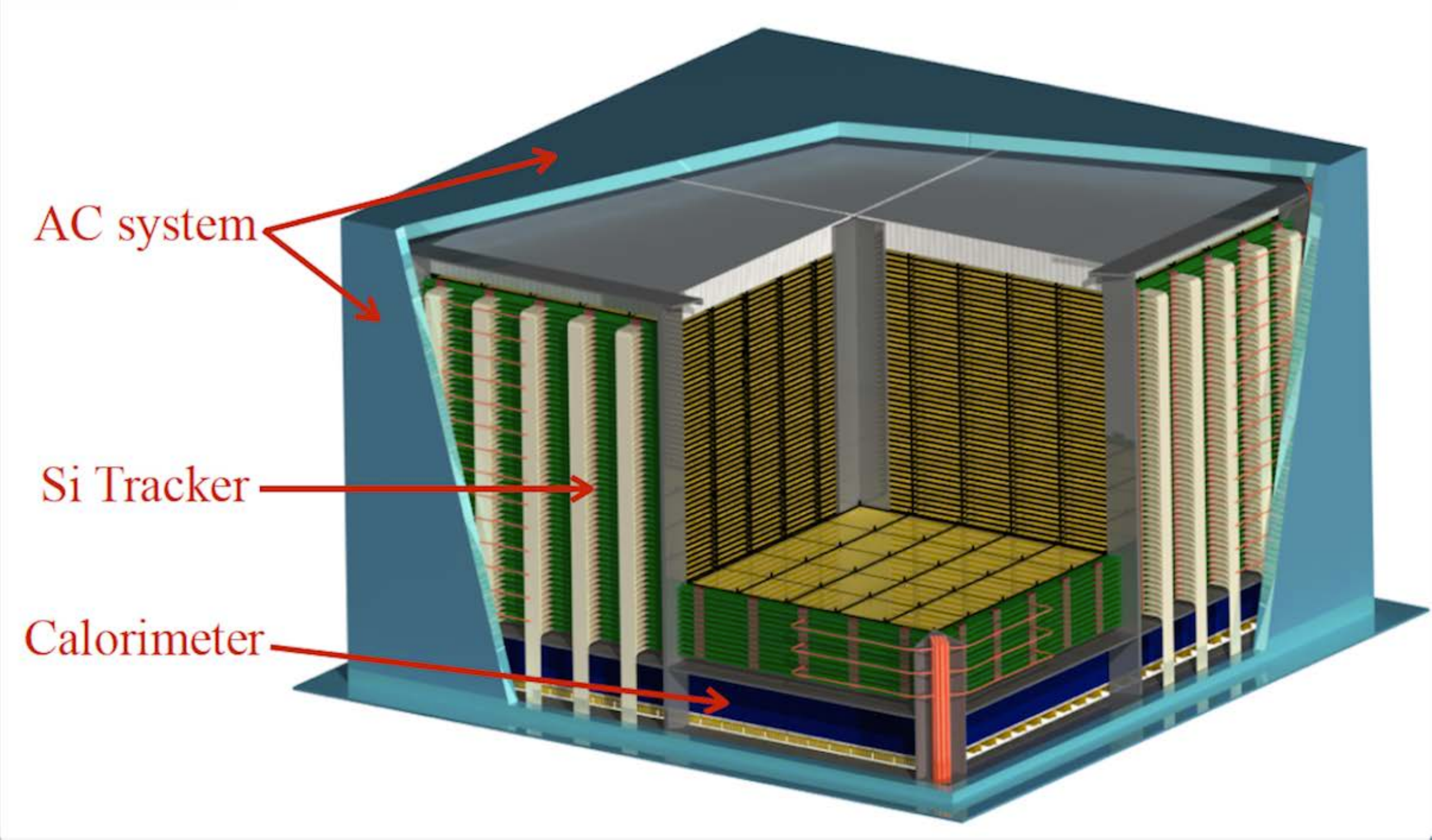}
\end{minipage}
\begin{minipage}{0.4\linewidth}
\includegraphics[scale=0.21]{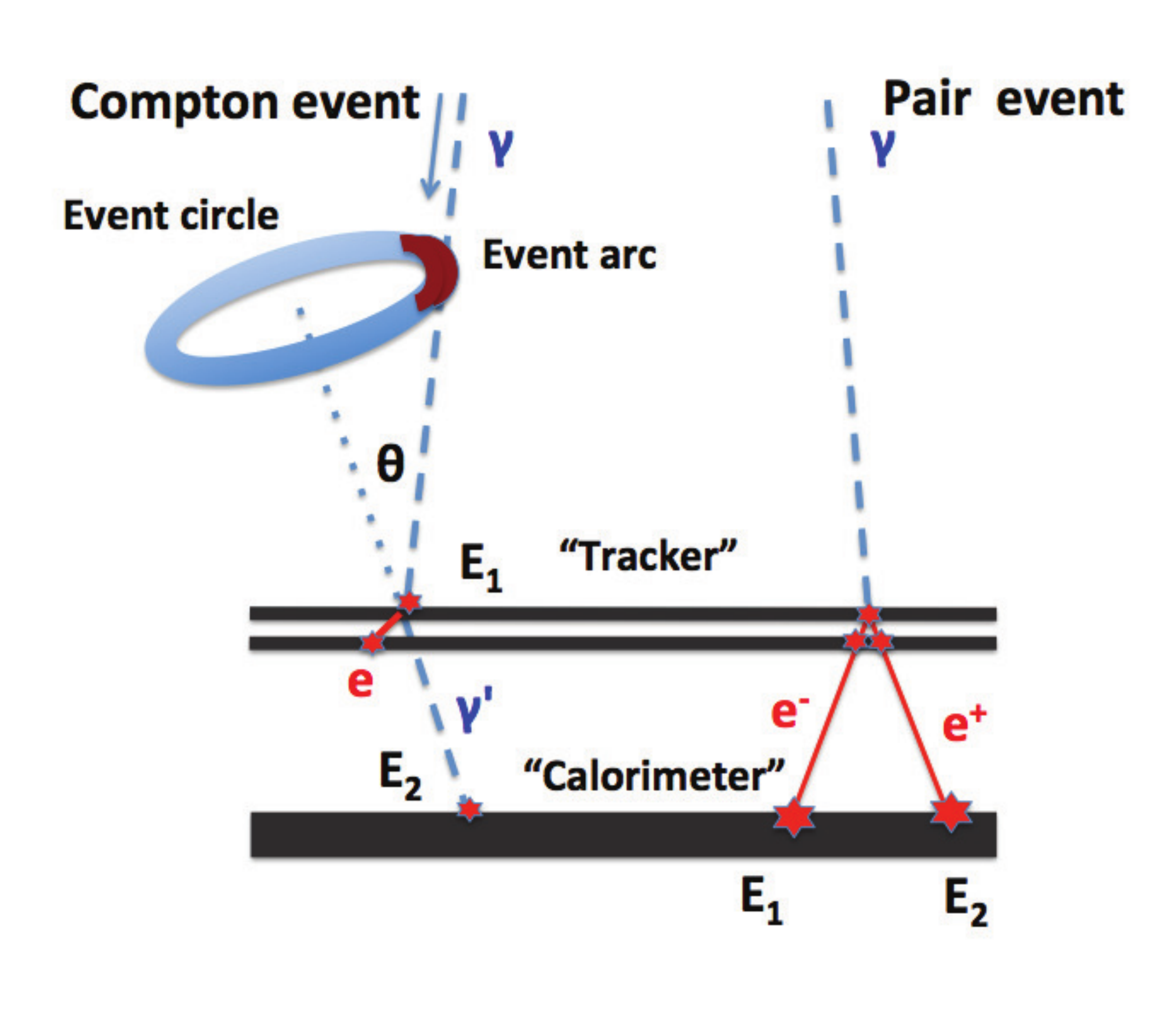}
\end{minipage}
\\
\end{tabular}
\end{center}
 \hspace{0.9cm} (a) \hspace{4.2cm} (b) \hspace{4.6cm} (c)
\caption 
{ \label{fig:ea}
(a) The e-ASTROGAM satellite under Ariane 6.2 fairing in upper position. (b) Overview of the e-ASTROGAM telescope with its three main detectors. (c) Representative  topologies for a Compton event and  a pair event in the telescope. Photon tracks are shown in pale blue, dashed, and electron and/or positron tracks in red, solid.} 
\end{figure} 

e-ASTROGAM is a proposal for an open gamma-ray observatory dedicated to the exploration of the Universe with unprecedented sensitivity in the energy range 0.15 -- 100~MeV, extending up to about 3 GeV. Many of the most-spectacular objects in the Universe have their peak emissivity in this energy band (e.g. gamma-ray bursts, blazars, pulsars, etc.), which is thus essential to study the physical properties of these objects. This energy range is also known to feature a characteristic spectral turn-over associated to hadronic emission from pion decay, which makes it paramount for the study of cosmic ray acceleration, by making it possible to distinguish hadronic and leptonic processes of nonthermal radiation. Moreover, this energy domain covers the crucial range of nuclear gamma-ray lines produced in various astrophysical sites by radioactive decay, nuclear collision, positron annihilation, or neutron capture.  

The core mission science of e-ASTROGAM addresses three major topics of modern astrophysics:
\begin{itemize} 
\item Processes at the heart of the extreme Universe: jet and outflow astrophysics (active galactic nuclei, gamma-ray bursts, compact binaries) and the link to new astronomies (gravitational waves, neutrinos, ultra-high energy cosmic rays);
\item Origin and impact of high-energy particles on the evolution of the Milky Way, from cosmic rays to antimatter;
\item Supernovae, nucleosynthesis and the chemical evolution of our Galaxy.
\end{itemize}
In addition, e-ASTROGAM is ready for many serendipitous discoveries (the ``unknown unknowns'') through its combination of wide field of view and much improved sensitivity compared to previous missions in this domain, measuring  in three years the spectral energy distributions of thousands of Galactic and extragalactic sources. e-ASTROGAM will also provide a groundbreaking capability for measuring gamma-ray polarization, thus giving access to a new observable that can provide valuable information on the geometry and emission processes of various high-energy sources. 

The e-ASTROGAM payload consists of a single telescope for the simultaneous detection of Compton and pair-producing gamma-ray events (see Figure~\ref{fig:ea}). It is made up of three detection system: a silicon Tracker in which the cosmic gamma rays undergo a Compton scattering or a pair conversion, a scintillation Calorimeter to absorb and measure the energy of the secondary particles, and an anticoincidence system to veto the prompt-reaction background induced by charged particles. The telescope has a size of 120$\times$120$\times$78 cm$^3$ and a mass of 1.2~tons (including maturity margins plus an additional margin of 20\% at system level). 

The e-ASTROGAM mission is presented at length in \cite[De Angelis et al. (2017)]{ea}. Here, we discuss the breakthrough capabilities of the proposed observatory to study the physics of both thermonuclear and core-collapse supernovae (SNe), as well as the process of cosmic-ray acceleration in SN shocks. 

\section{Thermonuclear supernovae}

\begin{figure}
\centering
\includegraphics[width=0.48\textwidth]{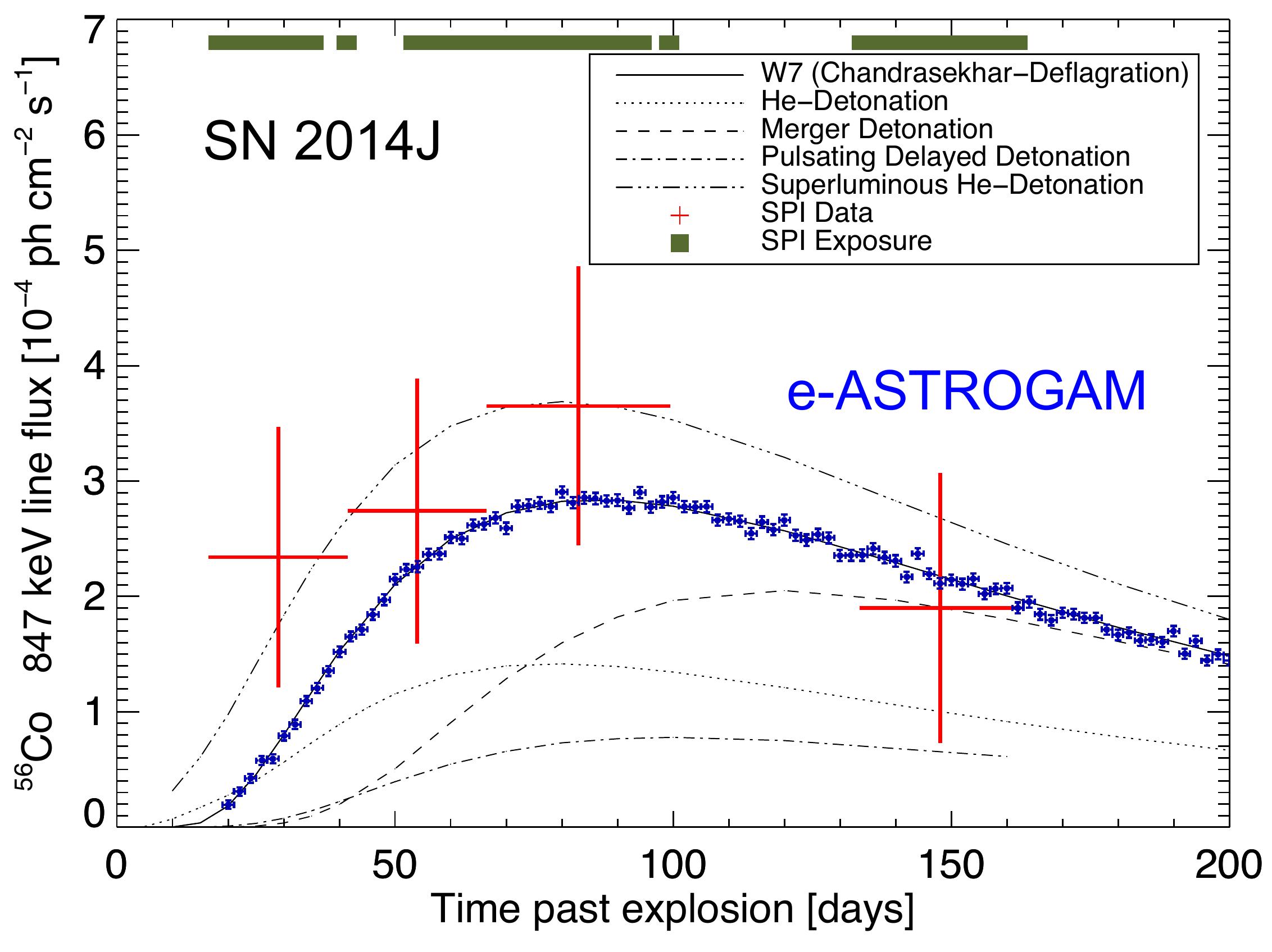}
\caption{Light curve of the 847 keV line from $^{56}$Co decay in SN~2014J. INTEGRAL data (adapted from Fig.~4 in \cite[Diehl et al. (2015)]{diehl15}; red data points) are compared to various models of Type Ia SN (\cite[The \& Burrows 2014]{the14}). A simulation of the e-ASTROGAM response to a time evolution of the 847 keV line such as in the W7 model (\cite[Nomoto et al. 1984]{nomoto84}) shows that the sensitivity improvement by e-ASTROGAM (blue points) will lead to a much better understanding of the SN progenitor system and explosion mechanism.}
\label{fig:sn2014j}
\end{figure}

SNe Ia are the outcome of a thermonuclear disruption of a white dwarf, either through centrally-ignited energy release from carbon fusion reactions, or from external triggering such as, e.g., collision/merging with a white dwarf companion star. But exactly how the ignition conditions are obtained, and on which white dwarfs, and more so how the thermonuclear runaway proceeds through the white dwarf and turns it into a variety of isotopes that are ejected, are all questions that are subject to considerable debate (e.g. \cite{hil00,hil13} and references therein). Such uncertainties are troublesome for cosmology since the use of SN Ia as standard candles depends on an empirical relationship between the shape and the maximum of the light curve (\cite[Phillips 1993]{phi93}), which is closely related to the mass of synthesized $^{56}$Ni. Factors like the progenitor evolution, ignition density, flame propagation, mixing during the burning, completeness of burning in outer, expanding regions, all lead to different amounts of synthesized $^{56}$Ni, which, however, can be measured directly through gamma-ray lines  (see contribution by Diehl, this volume).

The first detection of a thermonuclear SN in gamma rays was realized with SN~2014J and INTEGRAL's instruments (\cite[Churazov et al. 2014, 2015; Diehl et 2015]{chur14,chur15,diehl15}). The detected lines of $^{56}$Co (lifetime of $\sim$111 days) at 847 and 1238 keV provided good constraints on the mass of radioactive material and the expansion velocity of the ejecta. Moreover, possible signatures of the radioactive  $^{56}$Ni (mean lifetime $\sim$8.8 days) lines early-on have been reported  (\cite[Diehl et al. 2014; Isern et al. 2016]{die14,ise16}), suggesting either a surface explosion triggered from the companion or very special morphology of the runaway, in contrast to any conventional model. 

e-ASTROGAM will achieve a major gain in sensitivity compared to INTEGRAL for the main gamma-ray lines arising from $^{56}$Ni and $^{56}$Co decays. Thus with an expected sensitivity of $3.5 \times 10^{-6}$ ph~cm$^{-2}$~s$^{-1}$ for the 847~keV line in 1~Ms of integration time, which represents an improvement in sensitivity by a factor of about 70 compared to INTEGRAL/SPI, e-ASTROGAM will be able to detect events like SN~2014J up to a distance of 35 Mpc. In the corresponding volume, that includes, in particular, the Virgo cluster of galaxies, one can expect about 10 type Ia SN explosions in 3 years (nominal mission lifetime). As illustrated in Figure~\ref{fig:sn2014j}, e-ASTROGAM will provide much better data than we have now with INTEGRAL for SN~2014J from similarly nearby events. These data will allow us to probe the explosion mechanism in detail, and compare with astrophysical models for each event to better understand the progenitor system(s) and the thermonuclear explosion process. Moreover, the detected sample of thermonuclear SNe would provide a clean and fundamental test of the \cite[Phillips (1993)]{phi93} relation for a dominant population of Branch Normal type Ia SNe. The contribution of positrons to energy deposits from radioactivity will also be addressed by e-ASTROGAM, due to its excellent sensitivity for tracking the light curve of the 511 keV annihilation line.

\section{Core-collapse supernovae}

Similar to SN Ia, core collapse physics is also not well understood in terms of an astrophysical model (e.g., \cite[Woosley \& Janka 2005]{woo05}; \cite[Janka 2012]{jan12}). But these events are more common, being the end states of the evolution of massive stars, and are key to understanding the diversity of elements in the universe. Also here, deviations from spherical symmetry are the rule. The goal is to explain a tremendous variety of core collapse events, e.g. electron capture supernovae such as the Crab, clumpy explosions such as Cassiopeia A, collapsars that appear as gamma-ray burst sources and produce stellar mass black holes, superluminous supernovae that may be powered entirely differently by magnetar rotational energy, or pair instability supernovae that create huge amounts of radioactive $^{56}$Ni. 

Stellar rotation is known to exist but is complicated to track in its effects on stellar evolution, yet important for many of the above outcomes: nucleosynthesis, pre-supernova structure, core collapse. Measuring nucleosynthesis products such as $^{44}$Ti, $^{56}$Co, and $^{56}$Ni is one of the more direct ways to extract information on the inner processes triggering the explosion near the newly forming compact stellar remnant (e.g., \cite[Grefenstette et al. 2014, 2017]{gre14}) -- other observables are indirect, and mostly reflect interactions within the envelope, or with circumstellar, pre-explosively ejected, or ambient gas. 

With a gain in sensitivity for the $^{44}$Ti line at 1157~keV by a factor of 27 compared to INTEGRAL/SPI, e-ASTROGAM should detect the radioactive emission from $^{44}$Ti (half-life of $58.9 \pm 0.3$~yr) from most of the young (age $\lesssim$ 500 yr) SNRs in the Milky Way (e-ASTROGAM expected sensitivity for the 1157~keV line: $3.6 \times 10^{-6}$ ph~cm$^{-2}$~s$^{-1}$ in 1~Ms of integration time). The observatory thus will measure precisely the amount of $^{44}$Ti in the remnant of SN~1987A, which is currently disputed in the literature (\cite[Grebenev et al. 2012]{gre12}; \cite[Boggs et al. 2015]{bog15}; \cite[Tsygankov et al. 2016]{tsy16}). These observations will give new insights on the physical conditions of nucleosynthesis in the innermost layers of a supernova explosion and the dynamics of core collapse near the mass cut.

e-ASTROGAM will also detect the signatures of $^{56}$Ni and $^{56}$Co decay from several core-collapse SNe in nearby galaxies. Comparing $\gamma$-ray characteristics of different classes of core-collapse SNe, possibly including the pair instability SNe with their order of magnitude higher $^{56}$Ni production (e.g., \cite[Gal-Yam et al. 2009]{gal09}), will probe potentially large variations in their progenitors and offer a direct view of their central engines. Rare core collapse events are predicted to have gamma-ray line brightnesses orders of magnitude above typical supernovae: pair instability and magnetar-powered jet explosions will reveal much larger amounts of $^{56}$Ni. e-ASTROGAM will reach out to a larger part of the nearby universe to constrain the rate of such events, if not detect them.

\section{Cosmic-ray acceleration in supernova shocks}

Galactic cosmic rays are believed to be accelerated in expanding shock waves initiated by SN explosions, at least up to the ``knee'' energy of the cosmic-ray spectrum at $\sim$10$^{15}$eV (e.g., \cite[Axford 1994]{axf94}). But several fundamental questions remain partly unanswered, in particular on the injection of particles in the acceleration process, the efficiency of the process as a function of time after explosion, and the maximum energy achieved by accelerated particles in SN shocks.  

Here also, gamma-ray observations are crucial. Thus, spatially and spectrally resolved observations of a handful of SNRs with {\it Fermi}-LAT have revealed important spectral modifications with the age of the system: young SNRs tend to be harder and fainter at GeV energies than older (\cite[Acero et al. 2016]{ace16}). The performance of e-ASTROGAM will open the way for spectral imaging of a score of SNRs, spanning ages from $10^3$ to $10^5$ years. The 0.15~MeV~--~3~GeV gamma-ray band covered by e-ASTROGAM is essential to separate the emission from relativistic electrons and nuclei above 100 MeV, so the new data can constrain how electrons and protons are differentially injected into the shock, how large and sometimes highly intermittent magnetic fields build up near the shock, and how the acceleration efficiency and the total cosmic-ray content of a remnant evolves as the shockwave slows down. Moreover, the bremsstrahlung emitting electrons seen with e-ASTROGAM have energies close to the radio synchrotron emitting ones, and lower than those seen in synchrotron X-rays, thus permitting tomographic reconstruction of the magnetic field and electron distributions inside the remnant. 

Older remnants often interact with molecular clouds that provide target gas for cosmic rays escaping the remnant (see, e.g., \cite[Uchiyama et al. 2012]{uch12}). Resolving the diffuse pion emission produced in those clouds against the bright Galactic background is essential to probe the cosmic-ray spectra that are actually injected into the interstellar medium. Imaging the remnant and shocked clouds both require an angular resolution better than 0.2$^{\circ}$ around one GeV and a sensitivity below $10^{-11}$ erg cm$^{-2}$ s$^{-1}$ above 50 MeV that e-ASTROGAM will achieve. 

The instrument may also detect the gamma-ray line emission between 0.3 and 10~MeV from nuclear excitation caused by low-energy (i.e. sub-GeV) cosmic-ray nuclei, opening a new and unique way to remotely measure the flux and elemental composition of these elusive particles. This long awaited detection (e.g. \cite[Ramaty et al. 1979]{ram79}) will be essential to better understand the role of low-energy cosmic rays in the Galactic ecosystem. 

\section{Recent supernova history and Galactic Pevatrons}

\begin{figure}
\centering
\includegraphics[width=0.65\textwidth]{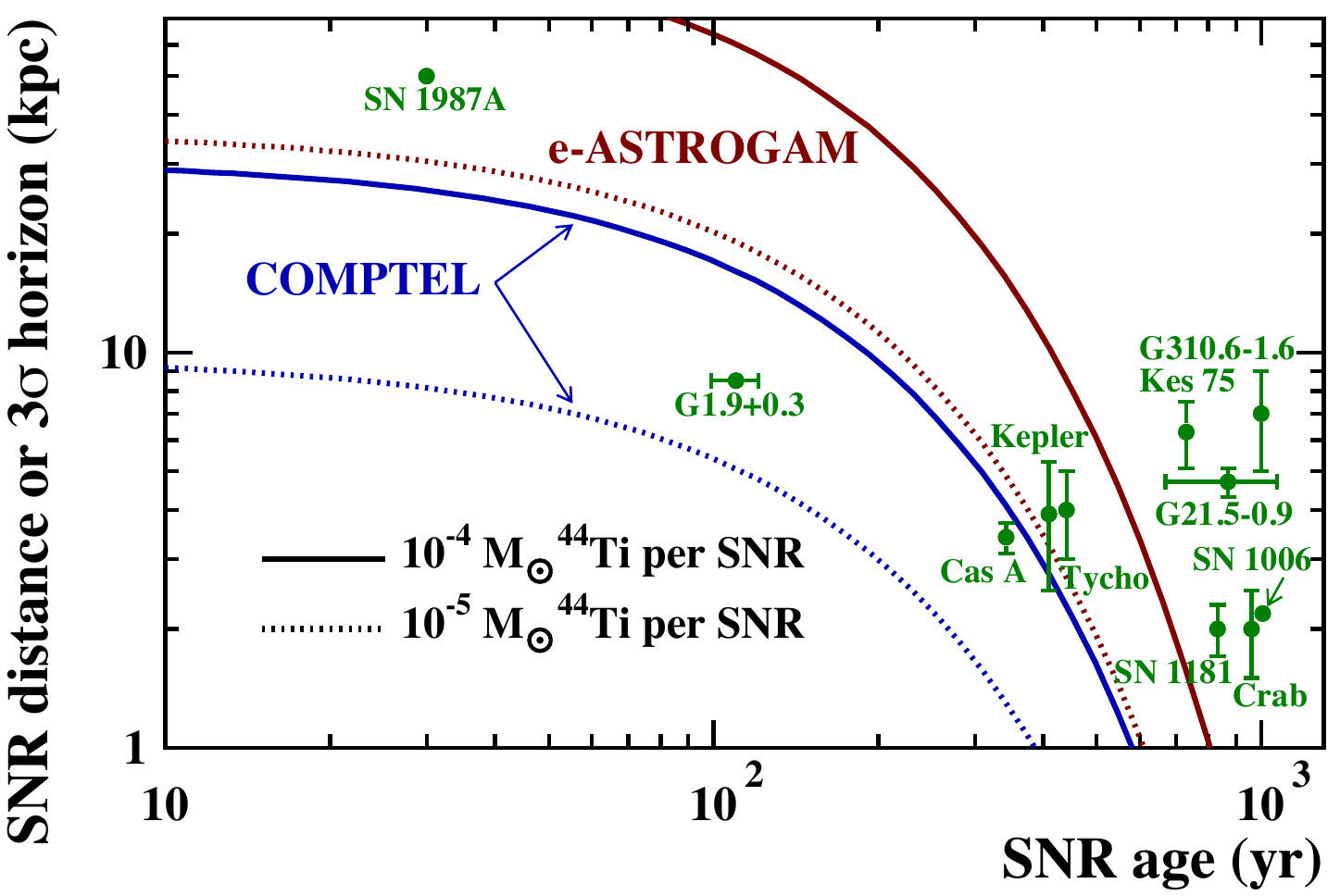}
\caption{Horizon of detectability of the $^{44}$Ti line at 1157~keV as a function of SNR age for CGRO/COMPTEL (blue lines) and e-ASTROGAM (red lines). The plotted sensitivities are for an effective source exposure of 1 year (COMPTEL: $9.0 \times 10^{-6}$ ph~cm$^{-2}$~s$^{-1}$; e-ASTROGAM: $6.4 \times 10^{-7}$ ph~cm$^{-2}$~s$^{-1}$), assuming two different yields of $^{44}$Ti production per SNR: $10^{-5}~M_\odot$ (typical of SNe~Ia; dotted lines) and $10^{-4}~M_\odot$ (Cas~A-like events; solid lines). Data for the age and distance of the known, young SNRs are from \cite[Tsygankov et al. (2016)]{tsy16}.}
\label{fig:ti44}
\end{figure}

Using the gamma-ray line emission from $^{44}$Ti decay, e-ASTROGAM is expected to uncover about 10 new, young SNRs in the Galaxy presently hidden in highly obscured clouds (see Figure~\ref{fig:ti44}), as well as the youngest SNRs in the LMC. Among the youngest Galactic SNRs, only Cassiopeia A has been firmly detected in $^{44}$Ti surveys carried out up to now (\cite[Tsygankov et al. 2016]{tsy16}), which is somewhat surprising in view of an otherwise inferred rate of one core-collapse SN every 50 years in our Galaxy (\cite[The et al.2006]{the06}; \cite[Dufour \& Kaspi 2013]{duf13}). In particular, e-ASTROGAM should be able to measure for the first time the amounts of $^{44}$Ti in the remnants of the Type Ia SNe G1.9+0.3, Tycho, and Kepler (Figure~\ref{fig:ti44}). 
 
e-ASTROGAM observations of young SNRs will be very important for cosmic-ray acceleration physics. The origin of PeV (1 PeV = $10^{15}$~eV) cosmic rays is still uncertain, although young SNRs are generally considered as the most likely sources. However, there is still no observational evidence for the acceleration of cosmic rays in SN shocks up to PeV energies. The most direct signature for this process would be the detection of a nonthermal gamma-ray spectrum extending up to about 100~TeV. Thus, the identification of potential new Pevatrons with e-ASTROGAM would be invaluable when the Cherenkov Telescope Array (CTA) will be fully operational.

\end{document}